\documentclass[conference]{IEEEtran}
\IEEEoverridecommandlockouts
\usepackage{cite}
 
\usepackage{amsmath,amssymb,amsfonts}
\usepackage{algorithmic}
\usepackage{graphicx}
\usepackage{textcomp}
\usepackage{xcolor}
\usepackage{booktabs}
\usepackage{array}
\usepackage{subcaption}
\usepackage{booktabs}
\usepackage{caption}
\usepackage[utf8]{inputenc}
\usepackage{url}
\usepackage{tabularx}
\usepackage{siunitx}
\usepackage{comment}

\def\BibTeX{{\rm B\kern-.05em{\sc i\kern-.025em b}\kern-.08em
    T\kern-.1667em\lower.7ex\hbox{E}\kern-.125emX}}
    
\title{Hate speech and hate crimes: a data-driven study of evolving discourse around marginalized groups\\}

\author{\IEEEauthorblockN{
Malvina Bozhidarova\IEEEauthorrefmark{2}~\IEEEauthorrefmark{8},
Jonathn Chang\IEEEauthorrefmark{3}~\IEEEauthorrefmark{8},
Aaishah Ale-rasool\IEEEauthorrefmark{1}~\IEEEauthorrefmark{9},
Yuxiang Liu\IEEEauthorrefmark{3}~\IEEEauthorrefmark{9},
Chongyao Ma\IEEEauthorrefmark{4}~\IEEEauthorrefmark{9},\\
Andrea L. Bertozzi\IEEEauthorrefmark{3},
P. Jeffrey Brantingham\IEEEauthorrefmark{5},
Junyuan Lin\IEEEauthorrefmark{6},
Sanjukta Krishnagopal\IEEEauthorrefmark{3}~\IEEEauthorrefmark{7}}\\

\IEEEauthorblockA{\IEEEauthorrefmark{1}\textit{Department of Mathematics}, \textit{University of Texas at Dallas}, Richardson, TX, USA}
\IEEEauthorblockA{\IEEEauthorrefmark{2}\textit{Department of Mathematics}, \textit{University of Nottingham}, Nottingham, UK}
\IEEEauthorblockA{\IEEEauthorrefmark{3}\textit{Department of Mathematics}, \textit{University of California, Los Angeles}, Los Angeles, CA, USA}
\IEEEauthorblockA{\IEEEauthorrefmark{4}\textit{Department of Mathematics}, \textit{Colby College}, Waterville, ME, USA}
\IEEEauthorblockA{\IEEEauthorrefmark{5}\textit{Department of Anthropology}, \textit{University of California, Los Angeles}, Los Angeles, CA, USA}
\IEEEauthorblockA{\IEEEauthorrefmark{6}\textit{Department of Mathematics}, \textit{Loyola Marymount University}, Los Angeles, CA, USA}
\IEEEauthorblockA{\IEEEauthorrefmark{7}\textit{Berkeley Artificial Intelligence Laboratory}, \textit{University of California, Berkeley}, Berkeley, CA, USA}\\
\IEEEauthorblockA{\IEEEauthorrefmark{8} Equal contribution \IEEEauthorrefmark{9} Equal contribution. \IEEEauthorrefmark{7}Corresponding author: sanjukta@math.ucla.edu}

\thanks{This material is based upon work supported by the National Science Foundation grants DMS-2027277 and DMS-2318817 and AFOSR-MURI grant FA9550-22-1-0380. MB was supported by the European Unions Horizon 2020 research and innovation programme under the Marie Skłodowska-Curie grant agreement No 777826 (NoMADS). Any opinions, findings, and conclusions expressed in this material are those of the authors.}
}

\begin{document}
\maketitle
\thispagestyle{plain}
\pagestyle{plain}

\begin{abstract}
This study explores the dynamic relationship between online discourse, as observed in tweets, and physical hate crimes, focusing on marginalized groups. Leveraging natural language processing techniques, including keyword extraction and topic modeling, we analyze the evolution of online discourse after events affecting these groups. Examining sentiment and polarizing tweets, we establish correlations with hate crimes in Black and LGBTQ+ communities. Using a knowledge graph, we connect tweets, users, topics, and hate crimes, enabling network analyses. Our findings reveal divergent patterns in the evolution of user communities for Black and LGBTQ+ groups, with notable differences in sentiment among influential users. This analysis sheds light on distinctive online discourse patterns and emphasizes the need to monitor hate speech to prevent hate crimes, especially following significant events impacting marginalized communities.
\end{abstract}


\section{Introduction}

In the wake of escalating racial tensions and recent surges in xenophobia, hate crime directed at marginalized communities has re-emerged as a significant problem in many settings \cite{Kaplan2023}. There is widespread concern that social media plays a significant role in promoting hateful behavior \cite{fanning}. While hate speech is a general challenge with wide relevance to society, its effects are particularly severe following major social disruptions such as the murder of George Floyd or the outbreak of the COVID-19 pandemic \cite{pandemicchange}.

Prior research has examined the complex interplay between online discourse and offline actions \cite{Gallacher2021,Bernroider_Harindranath_Kamel_2022}. Müller and Schwarz \cite{müller} compared occurrences of Muslim-related comments on Twitter to the frequency of anti-Muslim hate crimes. They found that tweets about Muslims by prominent online figures triggered increases in xenophobic tweets by others, cable news mentions of Muslims, and hate crimes against Muslims in the following days. Similar results on social media enabling the spread of extreme viewpoints that migrate to the physical world have been echoed in \cite{williams,fanning}. Conversely, Lupu et al. \cite{lupu2023offline} found that offline events, such as protests and elections, appear to trigger a rise in online hate speech. Lastly, Awan and Zempi \cite{10.1093/bjc/azv122} report on qualitative interviews suggesting that individuals targeted online are also frequently targeted offline.

The present work seeks to understand the broader contextual relationships between online hate speech and offline hate crime. We focus on a collection of tweets from California and examine their connections with hate crimes recorded from a publicly available dataset \cite{fbicrime}. The Twitter dataset is processed using keyword filtering and topic modeling to identify tweets related to Black, Asian, Hispanic, Jewish, and LGBTQ+ groups. We refer to tweets sorted by topic as ``topic groups" since a tweet's membership in a group is dependent on the content of the tweet, not on the characteristics of the individual who posted the tweet. As hate speech is not defined consistently, we use sentiment analysis to study the behavior of tweets with negative sentiment. We present statistics for negative sentiment tweets aimed at Black, Asian, LGBTQ+, Hispanic, and Jewish communities, but restrict ourselves to Black and LGBTQ+ groups for detailed analysis.

First, we perform time series analysis to identify significant temporal correlations between tweet topic groups and hate crimes for Black and LGBTQ+ groups and find that online and offline activity both peak around major societal events pertaining to these groups.

Following this, we draw on network analyses to study the social behavior among the Black and LGBTQ+ topic groups. Network models provide a robust framework to study time-evolving relationships that underlie various complex phenomena (e.g. social isolation and unemployment \cite{ROZER2020100}, friendship and violence among classmates \cite{WITTEK202034}) They can also provide concise graphical descriptions of data that lend easily to analysis of the most central nodes \cite{OPSAHL2010245}. 

In our research, we use a knowledge graph \cite{ibm_kg} to structure tweets and their multitude of attributes, leveraging the graphical structure to uncover connections between users and their discussions. We employ dynamic community detection \cite{Qanon} to compare the temporal evolution of communities of users associated with the Black and LGBTQ+ topic groups. In both topic groups, the emergence of new communities co-occurs with large-scale real-world events. In addition, the community of users in the Black topic group displays patterns of emergence and dissolution, while the community of users in the LGBTQ+ topic group tend to split over time but displays less dynamic activity. We also construct a user network, a complete graph connecting active users in each topic group weighted by the similarity of their tweets. Finally, we define `influence' as a way to measure user importance by combining network-based centrality metrics on the user network with follower count. We find that the most influential users in the Black topic group user network tweet more negatively and have more followers than the most central users in the LGBTQ+ topic group user network, highlighting important differences in their online behavior.

This paper is structured as follows. In Section \ref{datasets}, we introduce the raw data and filter it by keywords related to five underrepresented groups. In Section \ref{NLP}, we perform sentiment analysis and topic modeling on the filtered tweets and assign topics to these five targeted groups. In Section \ref{data analysis}, we conduct time series analysis on the Black and LGBTQ+ topic groups, comparing the frequencies of low-sentiment tweets and hate crimes related to these topic groups. In Section \ref{KG construction}, we construct the the knowledge graph and perform dynamic community detection to compare the evolution of communities on Black and LGBTQ+ topic group graphs. In Section \ref{user network}, we construct a weighted user network and compute centrality to compare the behavior of the most influential users among the Black and LGBTQ+ topic groups. In Section \ref{conclusion}, we summarize our findings and outline possible future directions.

\section{Datasets} \label{datasets}

\subsection{Twitter Dataset} \label{twitter}

To model online discourse, we collect a Twitter (now known as X) dataset consisting of more than 28 million tweets dating from March 11, 2020, to June 17, 2021. The tweets were collected using the Twitter API and were selected using the bounding box attribute to include those with user attributes or tweet locations tied to the greater Los Angeles area \cite{twitter2023}. Users may be associated with Los Angeles but active in another location. Thus, most tweets are concentrated in Los Angeles, but other locations in California are also observed in the dataset. We focus on Los Angeles due to its status as a large and diverse metropolitan city in the US, but our ensuing methodologies can be extended to other areas given appropriate data.

Each tweet in the dataset is represented as an object containing a unique ID number, the date of its creation, raw text information, and separate fields for the tweet's hashtags and mentions. The tweet object also includes specific information about the user who posted it: the user's number of followers at the time of posting the tweet, their unique ID number, and their screen name at the time of posting the tweet. In addition, 81\% of the tweets have the ``lang" attribute set to English, but this increases to 88\% after filtering by keywords. We do not filter tweets by language because we are interested in representing the cultural diversity of online discussion in California. 

We process the data by filtering the tweets by their textual information through keyword extraction. The keywords are selected to match the most polarizing tweets, so they are typically derogatory or empowering towards specific groups, including slang or racial slurs. We source the keywords from academic journals and online glossaries that have accumulated lists of words relevant to our groups of interest \cite{ijerph17197032,PFLAG2023}. Examples of these keywords are given in Table \ref{table:keywords}, and the full list is given in a GitHub repository (\ref{code and data}). We omit words that do not specifically target the five groups as well as generic keywords that are typically non-polarizing, such as ``racism", ``justice", and ``equality". We augment the keyword list by including plurals, removing spaces, and replacing spaces with hyphens. The pre-augmented list contains 162 items, while the augmented list contains 694. We also clean the tweets by making all characters lowercase and removing punctuation, emojis, and non-alphanumeric characters. After filtering the tweets by keeping only the tweets that contain keywords in our augmented list, the processed dataset contains 74639 tweets.

\begin{table}
  \centering
  \begin{tabular}{@{}ll@{}}
    \toprule
    \textbf{Group} & \textbf{Keyword (Phrases)}                                 \\
    \midrule
    Black          & `blm', `george floyd', `blackout tuesday', `say her name'  \\
    Asian          & `amplify asian voices', `chinese virus', `stop asian hate' \\
    LGBTQ+          & `homophobia', `trans lives matter', `drag queen'           \\
    Hispanic       & `illegal aliens', `border wall', `hispanic lives matter'   \\
    Jewish          & `jewish pride', `anti semitism', `nazism', `scapegoat' \\
    \bottomrule
  \end{tabular}
  \caption{Keyword Examples}
  \label{table:keywords}
\end{table}

\subsection{Hate Crimes Data} \label{hc data}
The hate crime dataset is taken from the FBI Crime Data Explorer, a digital repository under the FBI's Uniform Crime Reporting (UCR) program \cite{fbicrime}. This data contains approximately 220k incidents from 1991 to 2021, including reports from all U.S. counties that contribute to the FBI.

The FBI defines hate crimes as offenses ``motivated, in whole or in part, by an offender's bias against a race, gender, gender identity, religion, disability, sexual orientation, or ethnicity, and committed against people, property, or society" \cite{fbibias}. Due to the substantial evidence required to classify an incident as a hate crime and the lack of reporting from certain counties, the FBI's dataset has been criticized for underestimation \cite{müller}. Still, it remains the most comprehensive, publicly available source of information on hate crimes. Each recorded incident includes details such as the date, location type (e.g., household, street, park), offense type (e.g., vandalism, assault, intimidation), and the race and number of offenders if known. The FBI also identified 35 motivating biases (e.g., anti-Black, anti-Asian) and classified each incident according to established evidence of bias.

\begin{figure*}[h!]
    \centering

    \begin{subfigure}[b]{0.475\linewidth}
        \includegraphics[width=\linewidth]{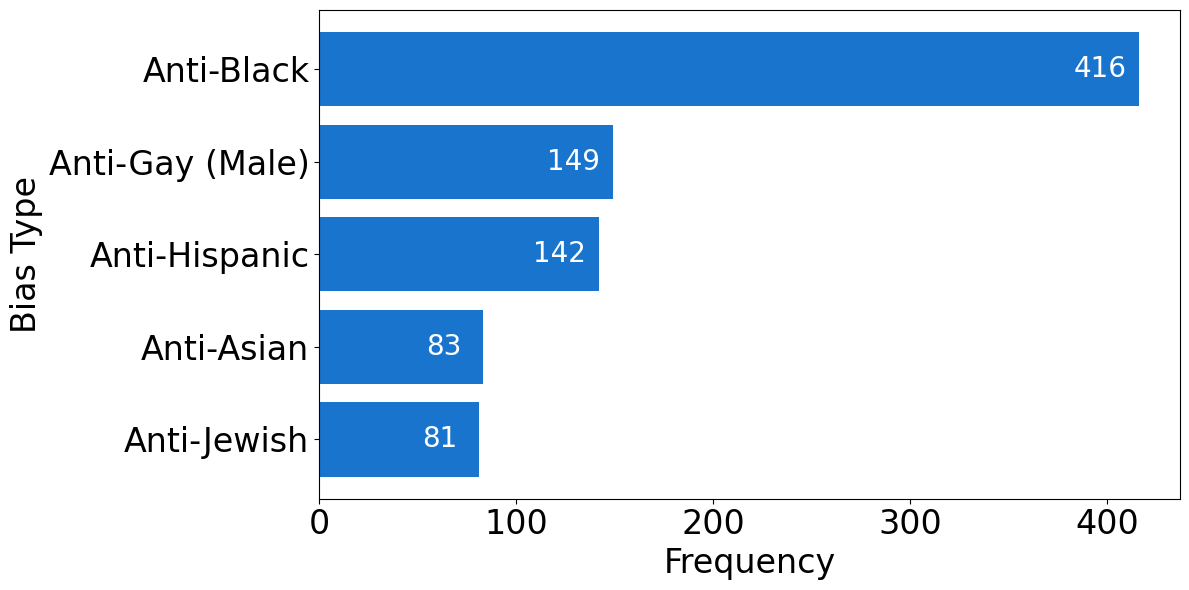}
        \caption{}
        \label{fig:top5biases}
    \end{subfigure}
    \hfill
    \begin{subfigure}[b]{0.475\linewidth}
        \includegraphics[width=\linewidth]{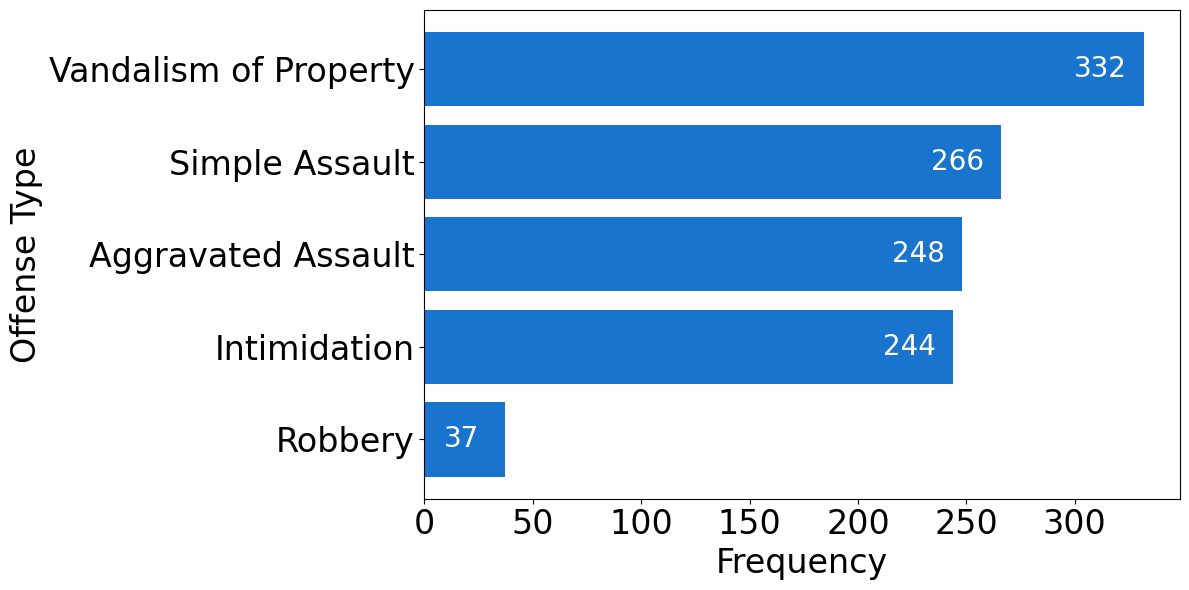}
        \caption{}
        \label{fig:top5crimetypes}
    \end{subfigure}

    \caption{Five most common motivating biases (left) and offense types (right) of hate crimes in California from 2020-03-11 to 2021-06-17.}
    \label{fig:top5biases and top5crimetypes}
\end{figure*}

In this work, we consider incidents that occurred in the state of California between March 11, 2020 and July 17, 2021, totaling 1192 hate crimes. We exclude crime data before March 11, 2020 and after July 17, 2021, in order to align the timestamps of hate crime and Twitter datasets. Within this timeframe, most hate crime incidents are racially motivated, followed by sexuality-based offenses. The primary offense types are property vandalism and simple assaults, followed by intimidation. Figure \ref{fig:top5biases and top5crimetypes} shows that the five most common bias types for hate crime events are anti-Black, anti-Gay (Male), anti-Hispanic, anti-Jewish, and anti-Asian.

\section{Natural Language Processing} \label{NLP}

\begin{figure*}[h!]
  \centering
  \includegraphics[width=\linewidth]{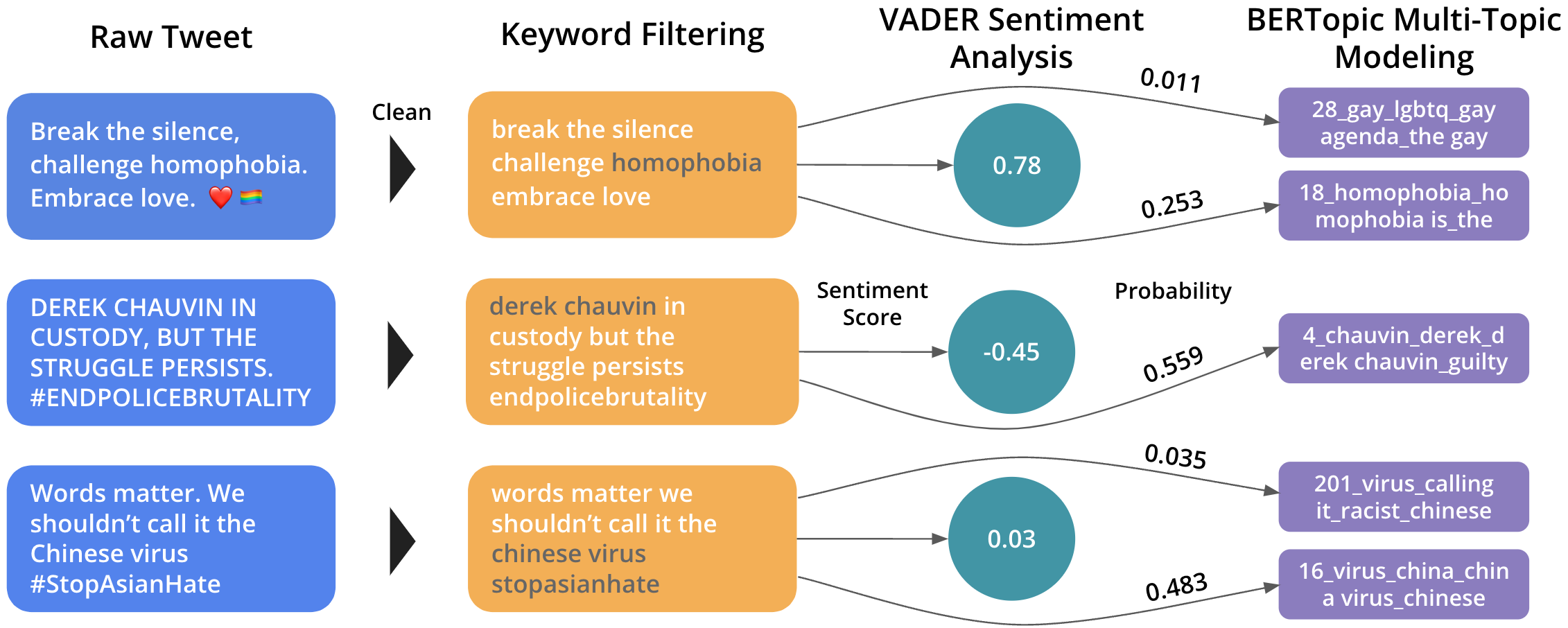}
  \caption{Natural Language Processing (NLP) Pipeline.}
  \label{fig:pipeline}
\end{figure*}

In this section, we discuss sentiment analysis and topic modeling, which are Natural Language Processing (NLP) methods used to extract information from the Twitter dataset. The NLP pipeline is outlined in Figure \ref{fig:pipeline}.

\subsection{Sentiment Analysis}\label{sentiment}
We use the Valence Aware Dictionary for sEntiment Reasoning (VADER) \cite{Hutto_Gilbert_2014} model for sentiment analysis. VADER is a state-of-the-art lexicon and rule-based sentiment analysis tool to extract social media sentiments. We select VADER for its efficiency in computation time as well as its interpretability and consistency in performance without the need of finetuning \cite{jain2023explaining}. Here, sentiment analysis is used to gauge public opinion, allowing us to identify, and analyze, the most polarizing tweets and users, providing insight into the most significant (and potentially harmful) trends within these online communities.

\subsection{Topic Modeling}\label{topic}

We leverage BERTopic, a widely used semantic-based topic modeling technique, to identify the most relevant topics in the filtered dataset. BERTopic first generates document embedding using the BERT model, then clusters these embeddings, and finally generates topic representations using a class-based TF-IDF procedure \cite{grootendorst2022bertopic}. This process categorizes the keyword-filtered tweets into a concise list of interpretable topics relevant to our groups of interest. BERTopic also serves as a noise filter by placing tweets with spam-like or noisy text, analogous to that generated by bots, into distinct topics of their own.

The topic model gives 270 topics, with the most prominent topic having 1540 tweets. We manually categorize topics into the following topic groups: 141 topics with 25467 tweets are related to a Black topic group, 16 topics with 3088 tweets to an LGBTQ+ topic group, 15 topics with 1846 tweets to an Asian topic group, 21 topics with 6585 tweets to a Jewish topic group, and 5 topics with 1000 tweets to a Hispanic topic group. The remaining topics are discarded. Topics that appear in more than one topic group are counted in each of the topic groups that they appear in, allowing us to identify connections between topic groups for the knowledge graph construction (see below). Around 38\% of the tweets, including a significant number of relevant tweets, are classified into an `irrelevant' or `miscellaneous' topic by BERTopic. To reduce the number of outliers, we also extract, from the clustering layer of BERTopic, the probability distribution over the topics for each tweet. Then, we move those tweets, originally classified into the miscellaneous topic, into one of the relevant topics if its probability for that topic is $\geq 0.01$. This allows us to classify each tweet into several topics, which we refer to as ``multi-topic modeling". We use the hyperparameters \texttt{min\_topic\_size} set to 30 to minimize the number of extraneous or repeated topics, as well as \texttt{top\_n\_words} and \texttt{n\_gram\_range} set to 15 and (1,3), respectively, to generate adequately sized topic names.

\begin{figure*}[t!]
  \centering
  \includegraphics[width=\textwidth]{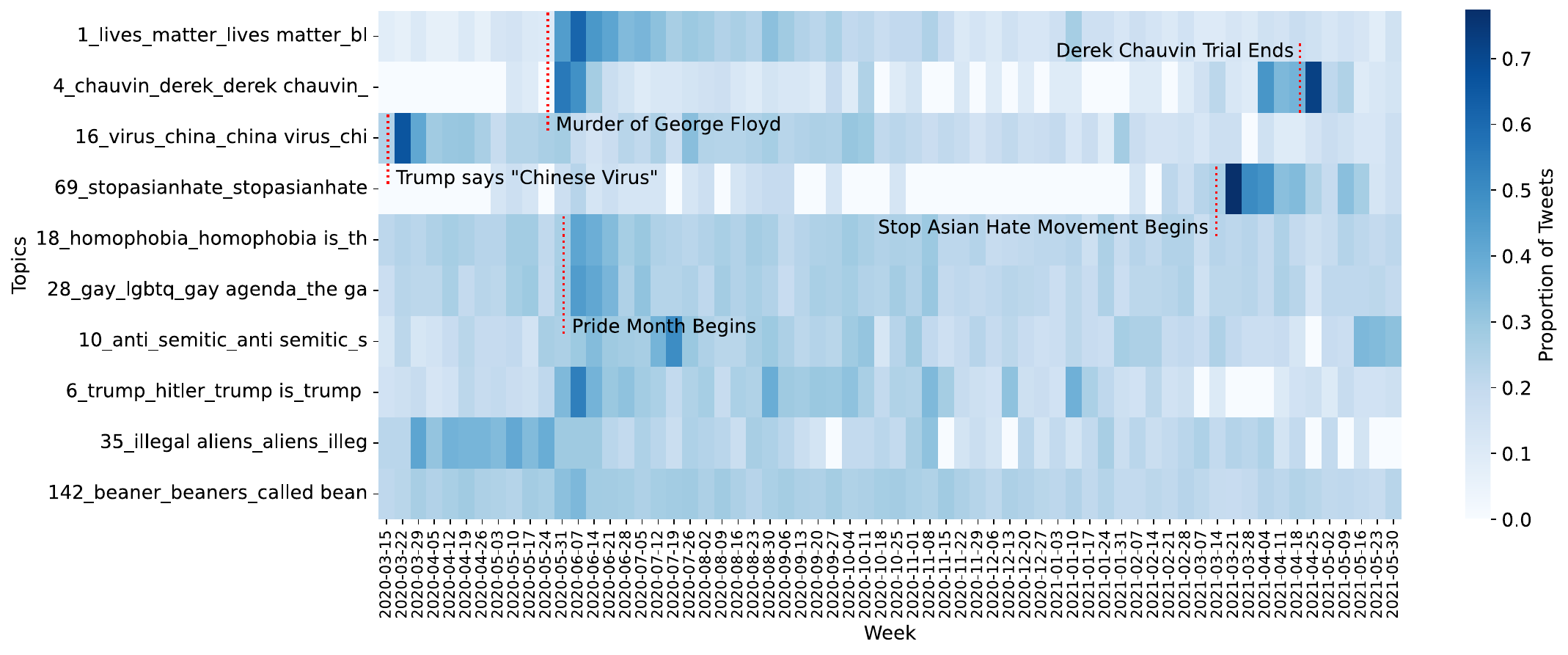}
  \caption{Proportion of tweets in topics from 2020-03 to 2021-06 weekly, annotated with relevant events. Two representative topics per group are selected randomly from each of the five topic groups: (top to bottom) Black, Asian, LGBTQ+, Jewish, and Hispanic. The shade of blue indicates the frequency of tweets from a specific topic (truncated) that occurred in a given week, relative to the total number of tweets in that topic. Real-world events relevant to a specific group are denoted by red dotted lines in the rows relevant to those groups.}
  \label{fig:topicvisual}
\end{figure*}

Figure \ref{fig:topicvisual} displays the weekly frequency of tweets in each topic as a fraction of the total tweets for two representative topics in each of our five topic groups. We see that tweet spikes in each topic group coincide with real-world events related to those topic groups. For example, the increase in the proportion of tweets with topics \texttt{1\_lives\_matter} and \texttt{4\_chauvin\_derek} corresponds to the Murder of George Floyd. Similarly, when Pride Month began in June 2020, the proportion of tweets with topics \texttt{18\_homophobia} and \texttt{28\_gay\_lgbtq} increased. This provides evidence that the topic groups identified by BERTopic are meaningful.

\section{Temporal data analysis} \label{data analysis}
    In this section, we examine the temporal correlation between online discourse and hate crimes targeting specific groups, and their relationship to real-world events. We focus on the Black and LGBTQ+ topic groups due to the availability of significantly larger amounts of data compared to other groups. We consider data from March 2020 to December 2020 and exclude data after December 31, 2020 due to an officially recognized discrepancy in collected crime statistics \cite{justice2021hatecrime} outside of this period that may yield misinformed correlations between discourse and crimes. 
    We extract tweets related to the LGBTQ+ and Black topic groups and study the temporal correlation between various quantities characterizing online activity (e.g., daily tweet count, average sentiment) and daily occurrences of hate crimes. We investigate trends in this correlation in the aftermath of major events such as the murder of George Floyd and Pride Month. To minimize noise in the daily data, we employ a 30-day rolling window for our time series analysis.

\begin{figure*}
    \centering

    \begin{subfigure}[b]{0.475\linewidth}
        \includegraphics[width=\linewidth]{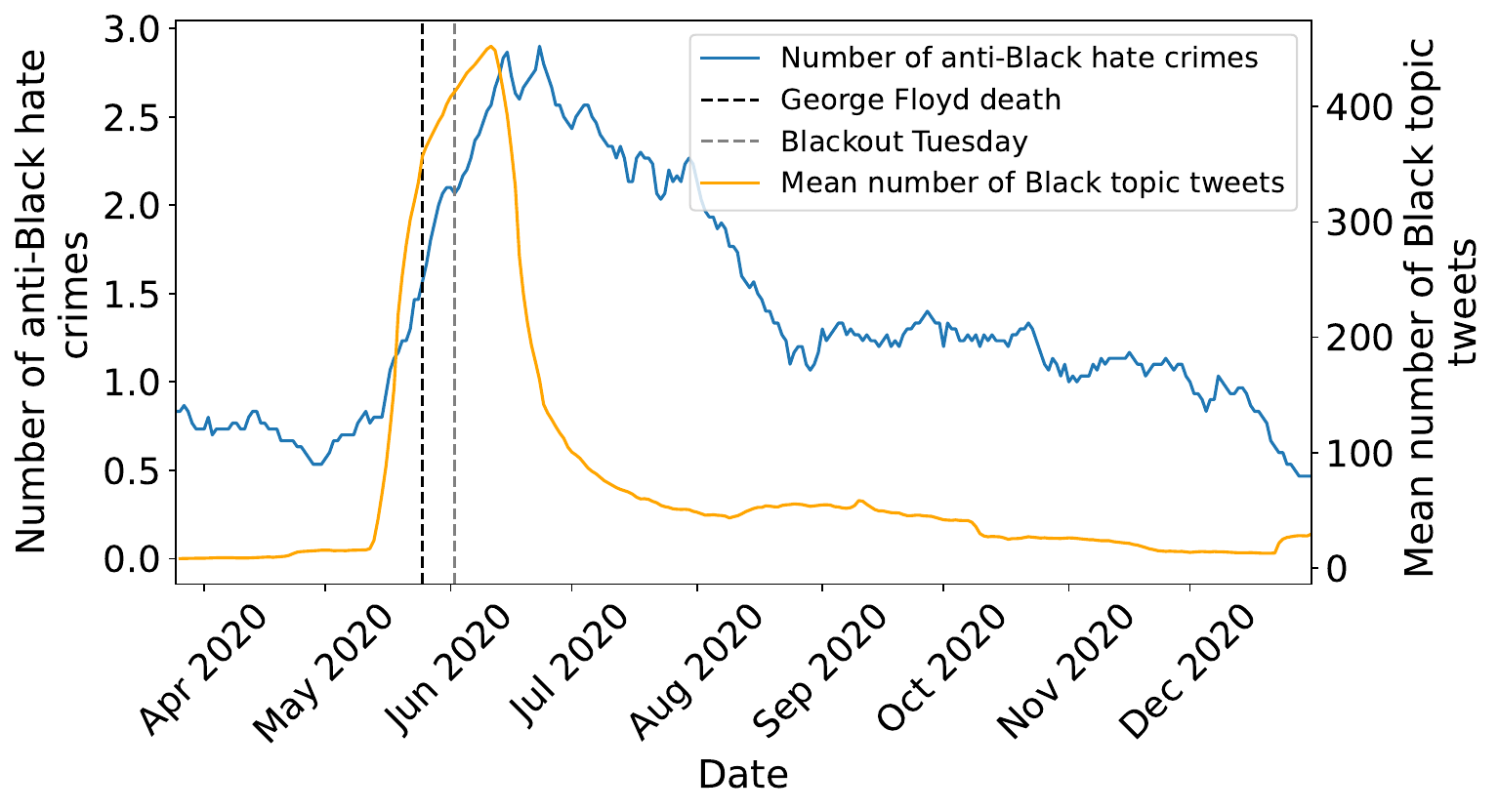}
        \caption{}
        \label{fig:numbers_black}
    \end{subfigure}
    \hfill
    \begin{subfigure}[b]{0.475\linewidth}
        \includegraphics[width=\linewidth]{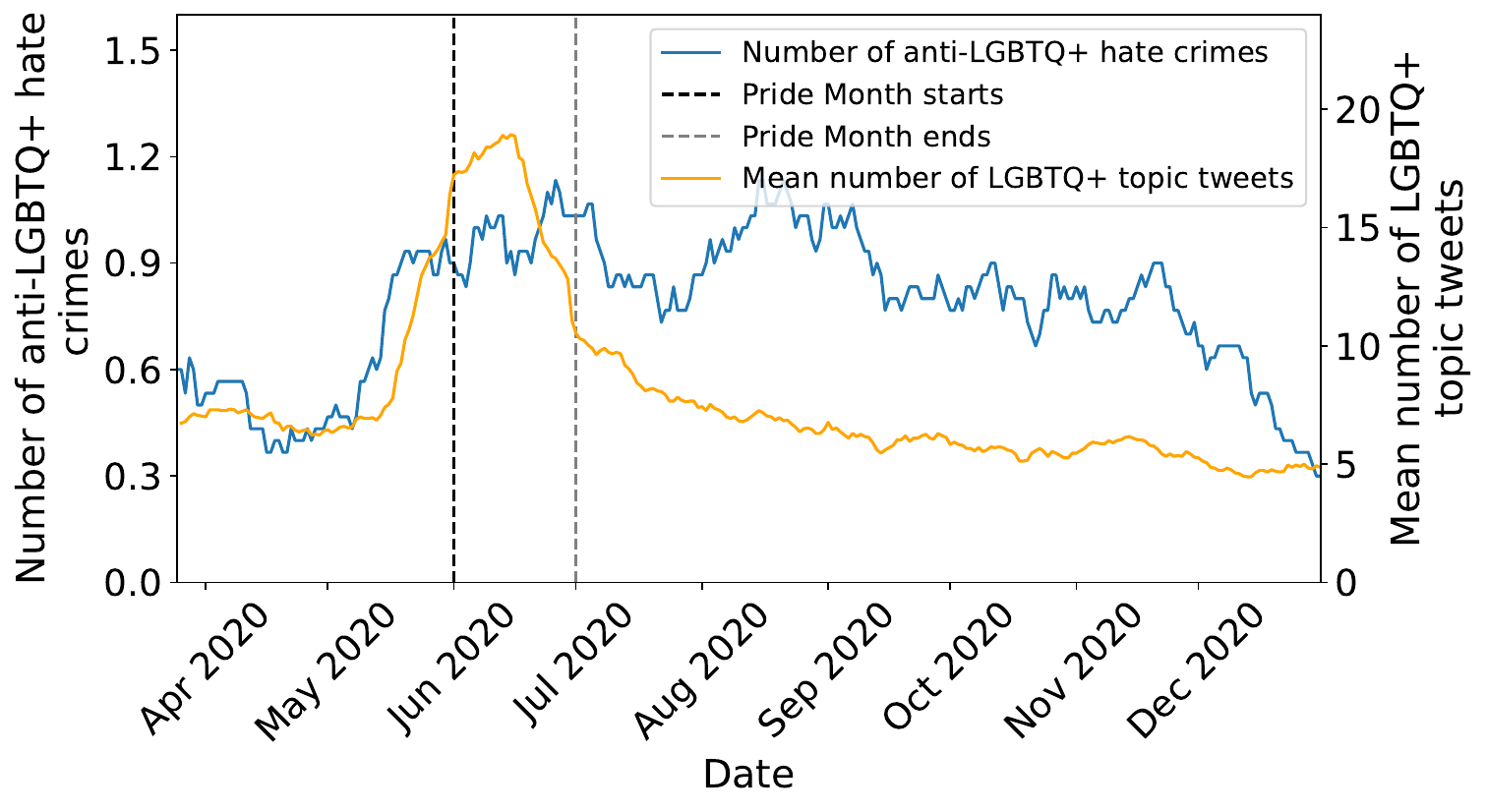}
        \caption{}
        \label{fig:numbers_lgbt}
    \end{subfigure}
\hfill
        \begin{subfigure}[b]{0.475\linewidth}
        \includegraphics[width=\linewidth]{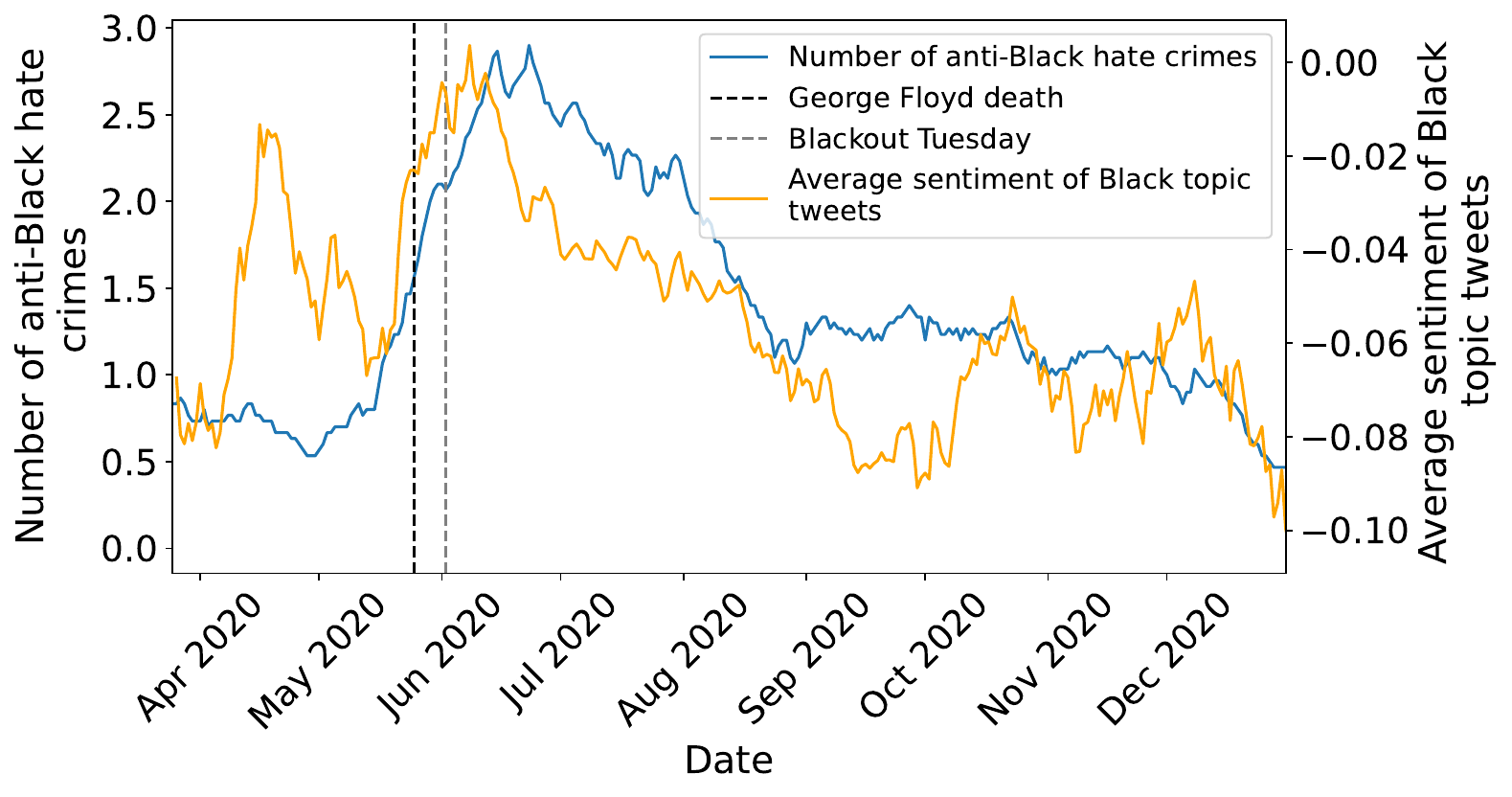}
        \caption{}
        \label{fig:sentiment_black}
    \end{subfigure}
    \hfill
    \begin{subfigure}[b]{0.475\linewidth}
        \includegraphics[width=\linewidth]{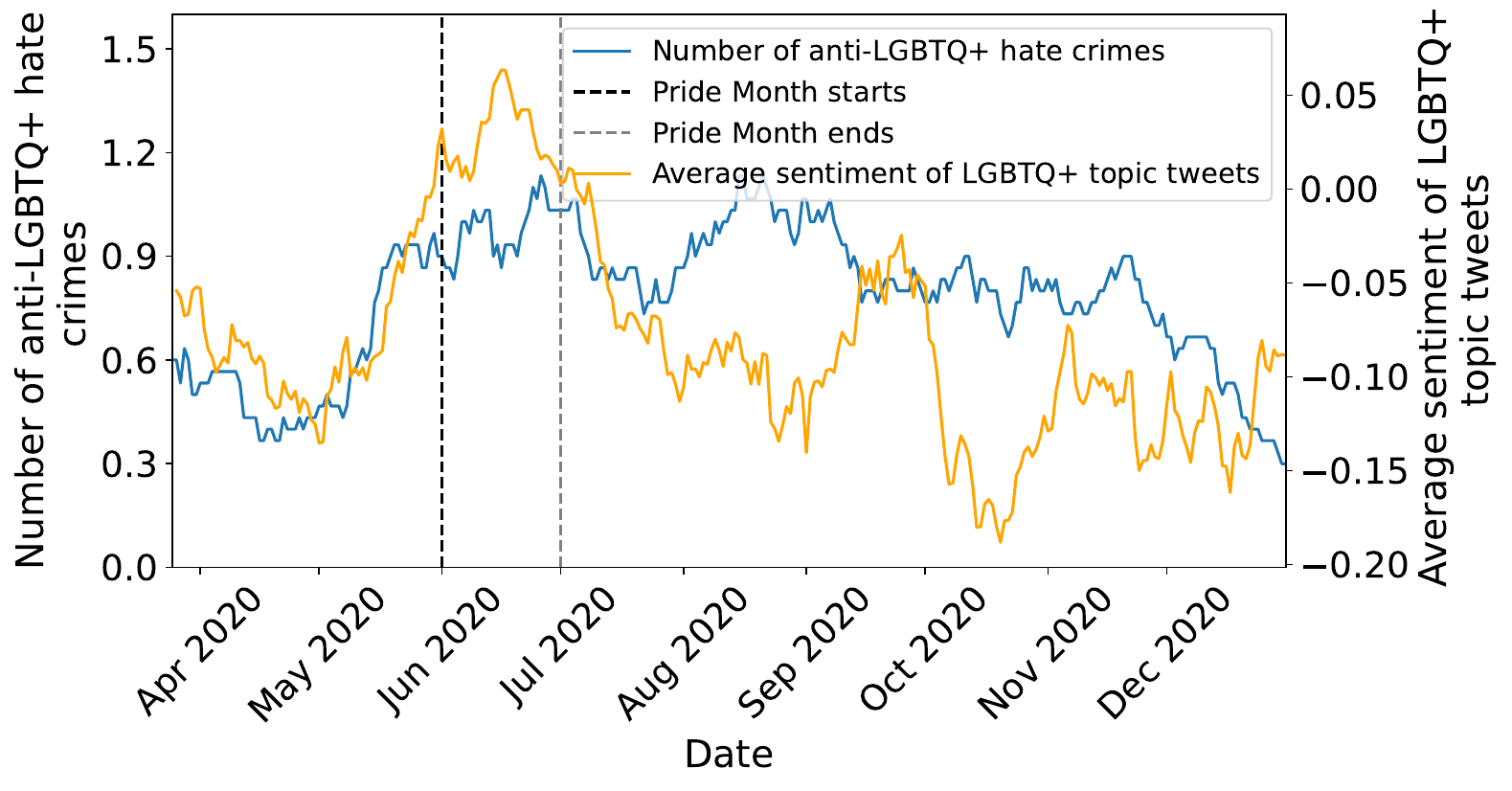}
        \caption{}
        \label{fig:sentiment_lgbt}
    \end{subfigure}
\hfill
        \begin{subfigure}[b]{0.475\linewidth}
        \includegraphics[width=\linewidth]{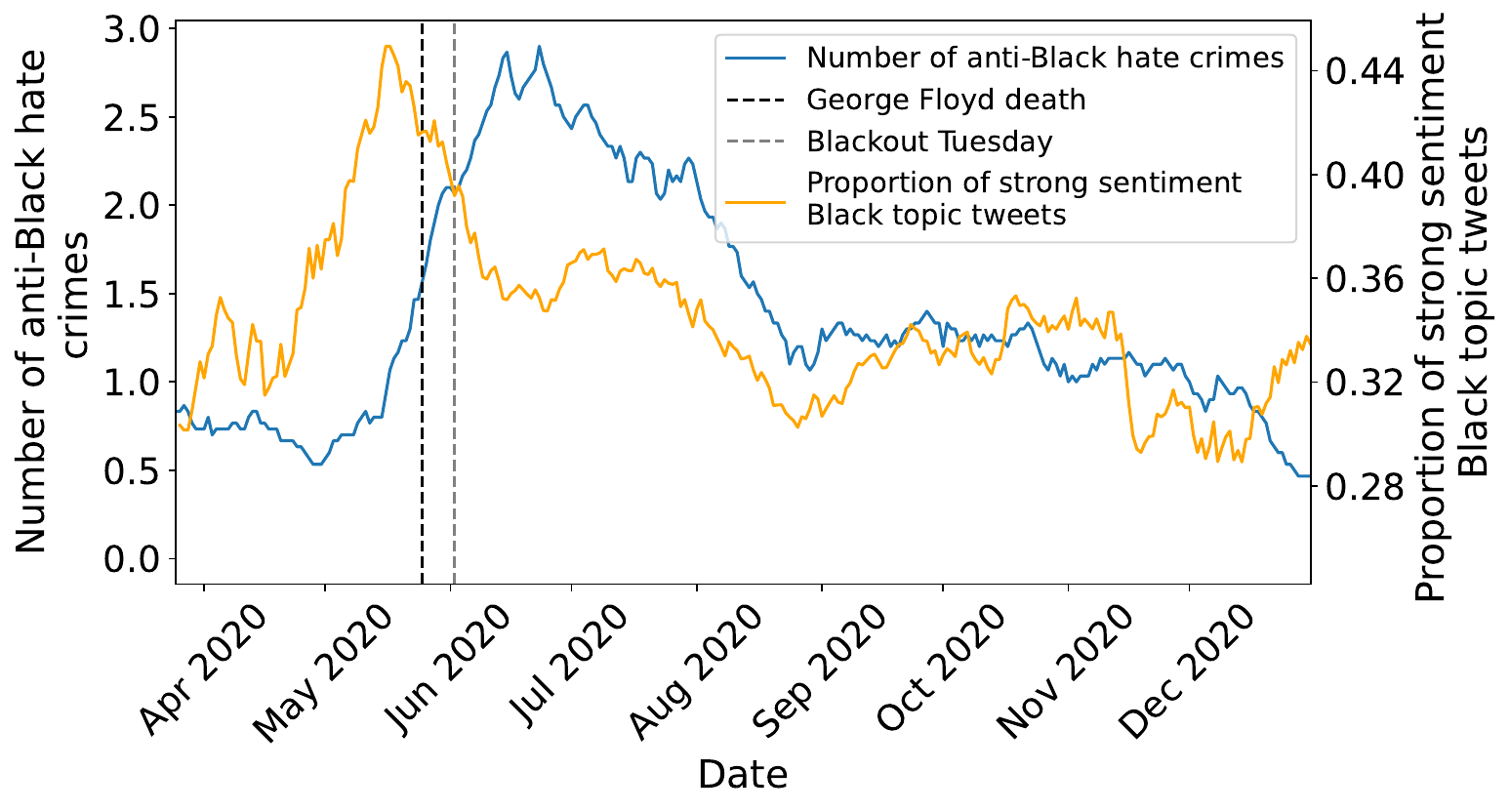}
        \caption{}
        \label{fig:strong_sentiment_black}
    \end{subfigure}
    \hfill
    \begin{subfigure}[b]{0.475\linewidth}
        \includegraphics[width=\linewidth]{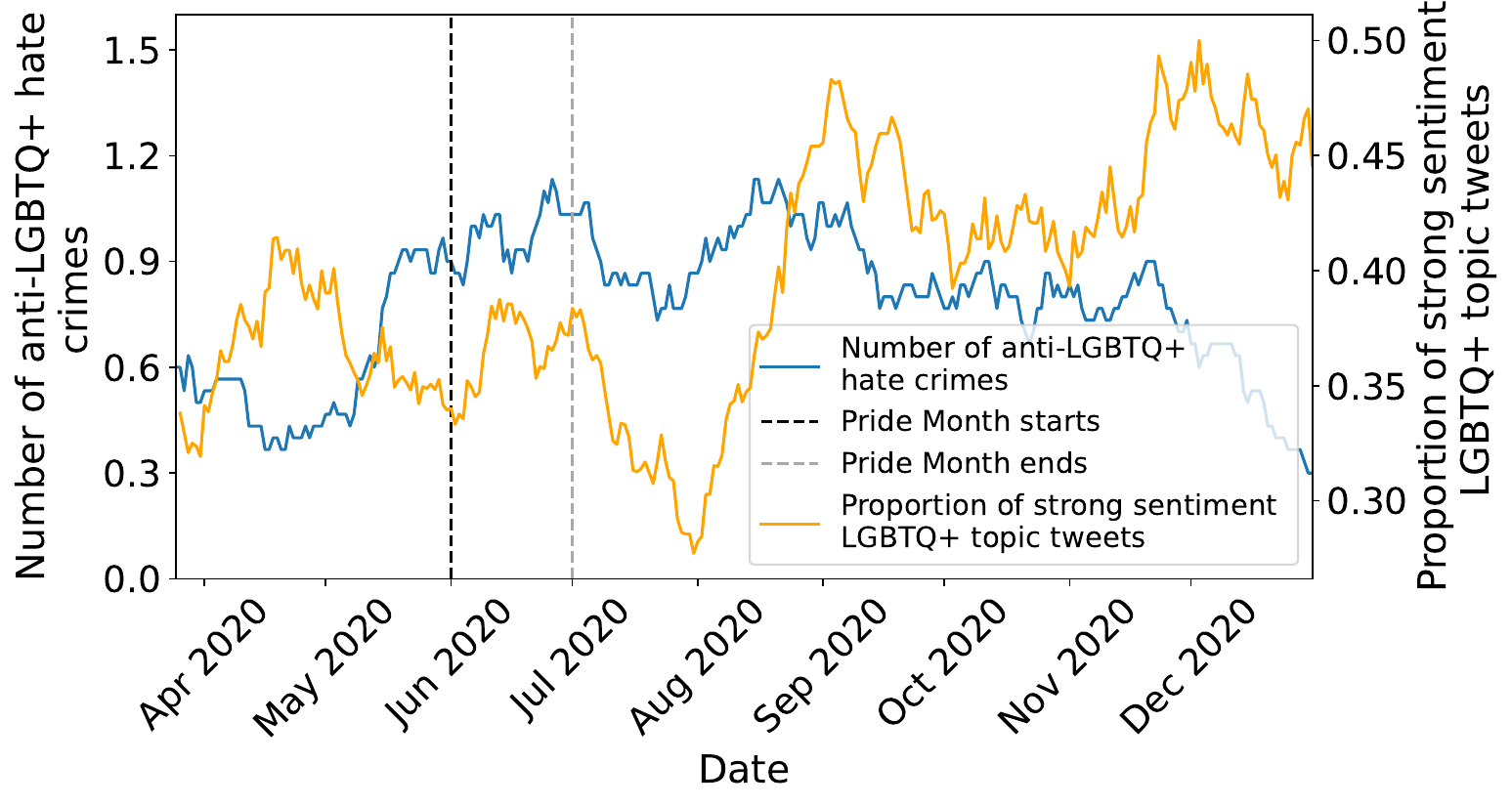}
        \caption{}
        \label{fig:strong_sentiment_lgbt}
    \end{subfigure}
\caption{The number of (left) anti-Black and (right) anti-LGBTQ+ hate crimes versus (a,b) the number, (c,d) the average sentiment, and (e,f) the proportion of ``strong'' sentiments, of all tweets in the respective topic groups. In each plot, the left y-axis corresponds to the blue curve, while the right y-axis corresponds to the orange curve.}
    \label{fig:number tweets vs number of hate crimes}
\end{figure*}

Figures \ref{fig:numbers_black} and \ref{fig:numbers_lgbt} illustrate the number of hate crimes and the total number of tweets in the Black and LGBTQ+ topic groups. In both figures, we see that the number of hate crimes and tweets increased after May 2020. In Figure \ref{fig:numbers_black}, there is a peak in anti-Black hate crimes following a peak in the Black topic group shortly after the murder of George Floyd, which received widespread public attention. Interestingly, while the frequency of tweets decreased rapidly over the subsequent month, the number of hate crimes stayed relatively high for up to six months. A similar peak in the LGBTQ+ topic group and hate crimes can be seen in Figure \ref{fig:numbers_lgbt} around Pride Month; while the number of tweets decreased following this, analogous to the Black topic group, anti-LGBTQ+ hate crimes are more consistent throughout 2020. 

Next, we explore the potential correlation between online sentiment and hate crimes. Figures \ref{fig:sentiment_black} and \ref{fig:sentiment_lgbt} show the average number of hate crimes compared to the average sentiment of tweets in the Black and LGBTQ+ topic groups, respectively. Sentiments take values between -1 and 1, and positive values are associated with positive sentiments whereas negative values are associated with negative sentiment. The absolute value of sentiment indicates how strong the sentiment is. Here we threshold the strength of a tweet to give a binary classifier; we consider a tweet to be `strong' if it has a sentiment with an absolute value greater than $0.5$. The results illustrate that the peak in the average sentiment for each group aligns with a corresponding major event---the murder of George Floyd and Pride Month, respectively. However, this peak is around 0 (neutral average sentiment) for the Black topic group, whereas it's 0.05 (positive average sentiment) for the LGBTQ+ groups, indicating the difference in the nature of the public events (death versus Pride week), and their effect on public sentiment. Additionally, the range of sentiment is larger (captured on the y-axis on the right) for LGBTQ+ groups.

Additionally, peak sentiment is aligned with peak hate crimes (particularly evident in the black topic group), which may initially seem counterintuitive as one might expect that higher mean sentiment in online discussions would correspond to fewer hate crimes. However, this can be explained by the unfolding of events after the murder of George Floyd. Immediately after, there were widespread protests and demonstrations against racial injustice and police brutality. These events galvanized a global movement advocating for racial equality and social justice and resulted in many people expressing positive sentiments online. 
We want to emphasize that we do not make any claims of causality here. 

Finally, we investigate a correlation between hate crimes and the strength of tweets in topic groups. Figures \ref{fig:strong_sentiment_black} and  \ref{fig:strong_sentiment_lgbt} illustrate the frequency of hate crimes targeting Black and LGBTQ+ topic groups alongside the proportion of tweets exhibiting a strong sentiment. We observe an interesting pattern where the rise in anti-Black hate crimes in June 2020 follows an increase in the proportion of strong sentiment tweets in May 2020. By contrast, an increase in anti-LGBTQ+ hate crimes corresponds to a lower proportion of strong tweets within that topic group. We also observe that the proportion of tweets that have strong sentiment is on average slgithtly higher in LGBTQ+ communities than black communities (their range is captured on the y-axis on the right). Moreover, the highest proportion of strong sentiment LGBTQ+ topic tweets aligns with the lowest proportion of strong Black topic tweets, suggesting that users may shift their focus between these topic groups. The results suggest that there is not one simple relationship between the strength of sentiments in online content and hate crime, but rather something that varies by group and context.


Next, we quantify the relationship between the temporal evolution of the above quantities. Table \ref{tab:spearman} presents the Spearman correlation coefficients $r_s$ \cite{spearman1907demonstration} and their associated $p$-values for each pair of time series in Figure \ref{fig:number tweets vs number of hate crimes}. The $r_s$ values show the strength and direction of monotonic relationships between each pair, while the $p$-values show the significance of the correlation. 
Table \ref{tab:spearman} shows that the correlation between the number of group-topic tweets and anti-group hate crimes differs for the Black and LGBTQ+ groups. While this correlation is not significantly different than $0$ ($p=0.773$) for the Black group, it is moderate and negative for the LGBTQ+ group. This suggests that changes in the number of Black-topic tweets do not appear to correlate with the occurrence of anti-Black hate crimes. In contrast, the negative correlation between the number of LGBTQ+ topic tweets and anti-LGBTQ+ hate crimes may be indicative of increased awareness, advocacy, education, and community support within online discussions, potentially contributing to a reduced occurrence of such hate crimes.
We also see that the average sentiment of tweets in both the Black and LGBTQ+ topic groups exhibits a negative correlation with the number of hate crimes towards the specific group, implying that sentiment of online discourse can affect public perspective and negative sentiment can lead to social environments that promote hate crimes.

Finally, we find that the correlation between the proportion of strong sentiment tweets and the number of hate crimes towards the targeted group display opposing patterns for the Black group (negative correlation) and the LGBTQ+ group (positive correlation). We postulate that the strong sentiment tweets in the Black group may lead to an increased level of awareness, concern, or activism, which may contribute to lower hate crime rates. Inversely, strong sentiment tweets in the LGBTQ+ group may lead to increased hate. Although speculative, these hypotheses could inform targeted strategies to detect and prevent the escalation of strong sentiment tweets into physical acts of hate.


\begin{table}[ht]
            \centering
            \begin{tabular}{@{}lcc@{}}
            \hline
            \textbf{Correlation with hate crimes of the specific group} & $\mathbf{r_s}$&$\mathbf{p}$\textbf{-value} \\ 
            \hline
            Number of Black topic tweets & 0.017& 0.773 \\
            
            Number of LGBTQ+ topic tweets & -0.545&5.60e-24\\
            \hline
            
            Average sentiment of Black topic tweets & -0.387& 7.69e-12 \\
            
            Average sentiment of LGBTQ+ topic tweets & -0.601& 4.70e-30 \\
            \hline
             Proportion of strong sentiment Black topic tweets & -0.352& 6.19e-10 \\
            
             Proportion of strong sentiment LGBTQ+ topic tweets& 0.463& 6.96e-17 \\
            \hline
            \end{tabular}
            \caption{The Spearman correlation coefficients and corresponding $p$-values between the number of anti-Black or anti-LGBTQ+ hate crimes and the number of tweets (top rows), the average sentiment of tweets (mid rows), and the proportion of strong sentiment tweets (bottom rows) in the Black or LGBTQ+ topic group respectively.}
            \label{tab:spearman}
        \end{table}

\section{Knowledge Graph} 
\begin{figure*}[ht]
  \centering
  \includegraphics[width=0.9\textwidth]{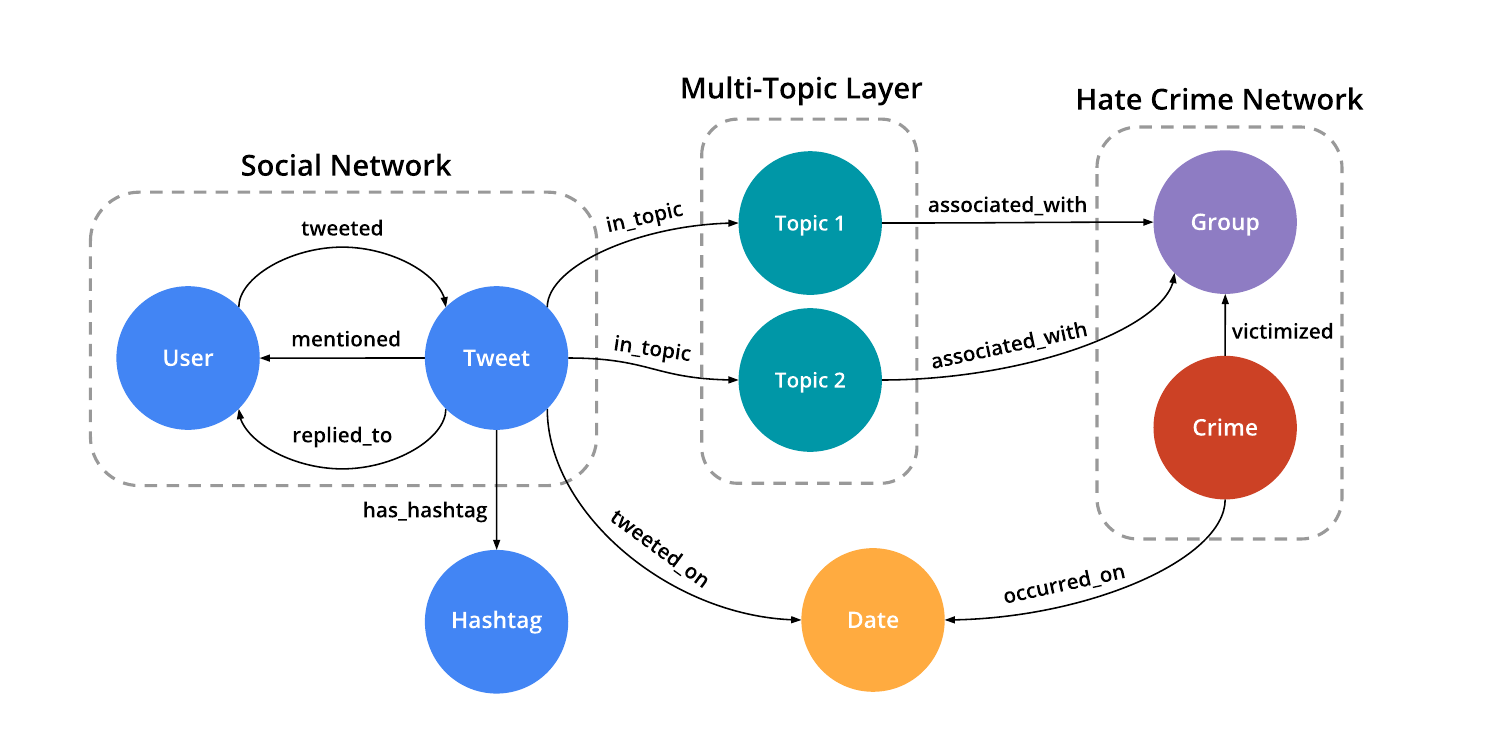}
  \caption{Ontology describing the entity and relation types in the KG. Blue entities are extracted from the Twitter dataset, teal entities form the Multi-Topic Layer connecting each tweet to its relevant topics, and the red and purple group and crime entities establish a network of real-world victims. Tweets and crimes are also connected by the date that they occurred.}
  \label{fig:ontology}
\end{figure*}

\label{KG construction}
A knowledge graph (KG) is a network consisting of nodes (real-world entities such as objects, events, and people) and edges (the relationships encoded between them) \cite{ibm_kg}. KGs provide a robust methodology to organize, structure, and analyze large quantities of data with complex relationships between different types of objects such as text data from various online platforms, and have been used successfully in a variety of data-driven research problems \cite{Zou_2020}.  A KG is particularly useful to identify complex patterns among topic groups and hate crimes that are not apparent from our time series analysis, by extracting subgraphs (e.g. extracting clusters of users and analyzing the sentiments of the most prevalent topics). We construct a KG using the semantic, sentiment, and topic information extracted from the tweets.

\subsection{KG Construction}
Our KG is a weighted and directed graph $G=(V, R, \Delta)$ where $V$ is the set of all nodes (consisting of seven different entity types), $R$ is the set of nine edge (relation) types, and $\Delta \subset V \times V \times R \times \mathbb{R}$ is the set of weighted triples. Each weighted triple $\alpha = (h,t,r,w)\in \Delta$ consists of a head and tail node $h, t \in V$, a relation type $r \in R$, and a weight $w\in [0,1]$ denoting the strength of the relation.
The entity and relation types are summarized in our ontology (Figure \ref{fig:ontology}). Five of these---tweeted, replied\_to, mentioned, has\_hashtag, tweeted\_on---are directly extracted from the Twitter dataset. For the in\_topic relation, we weight the edge using the probabilities calculated during topic modeling in Section \ref{NLP}. 

We filter for tweets with at least a 1\% likelihood of being in at least one of the 191 topics relevant to our five groups of interest, which leaves us with 57,142 tweets. 
The mean number of in\_topic relations is about four per tweet. The user and tweet entities, along with their relations, form a ``Twitter Social Network"---a subgraph encoding the complex interactions between users of different levels of influence. Incorporating the topic entities into the subgraph (which are generated by tweet text) allows us to perform analysis among tweets related to a certain topic group and examine the discourse over various subject matters in a group. Tweets and hate crimes are implicitly connected through the date entity. This enables temporal analysis of the KG through dynamic community detection.

We also introduce a group entity, which allows us to connect tweets about marginalized groups with the hate crimes targeting those groups. This is enabled by the multi-topic modeling, allowing us to categorize tweets into our desired topic group communities. The entity and relation counts for the completed KG are displayed in Table \ref{table:kgstats}.

    \begin{table}
    \centering
    \begin{tabular}{@{}llll@{}}
    \toprule
    \textbf{Entity}   & \textbf{Count}  & \textbf{Relation}         & \textbf{Count}  \\
    \midrule
    Tweet    & 57,142 & tweeted          & 57,142 \\
    User     & 22,065 & mentioned        & 604    \\
    Hashtags & 10,948 & replied\_to      & 2,533  \\
    Crime    & 738    & in\_topic        & 214,821\\
    Date     & 464    & has\_hashtag     & 55,127 \\
    Topic    & 191    & tweeted\_on      & 57,142 \\
    Group    & 5      & occurred\_on     & 738    \\
    Total    & 91,553 & victimized       & 738    \\
             &        & associated\_with & 198    \\
             &        & Total            & 389,043\\
    \bottomrule
    \end{tabular}
    \caption{KG Statistics}
    \label{table:kgstats}
    \end{table}

\subsection{Dynamic Community Detection} \label{community detection}

\begin{figure*}[h!]
    \centering
      \begin{subfigure}[b]{0.475\linewidth}
        \includegraphics[width=\linewidth]{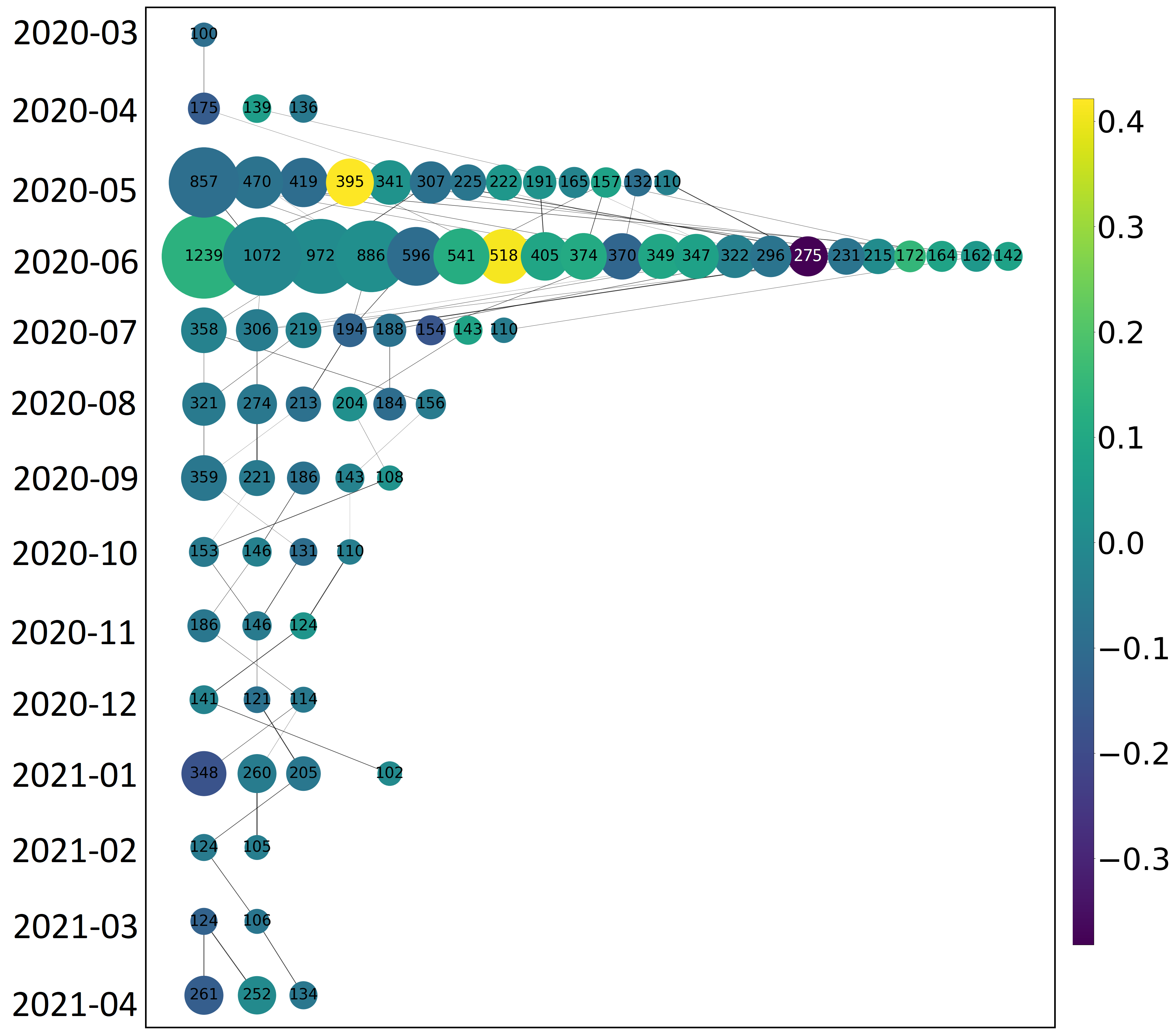}
        \caption{}
        \label{fig:community_users_black}
    \end{subfigure}
    \hfill
    \begin{subfigure}[b]{0.475\linewidth}
        \includegraphics[width=\linewidth]{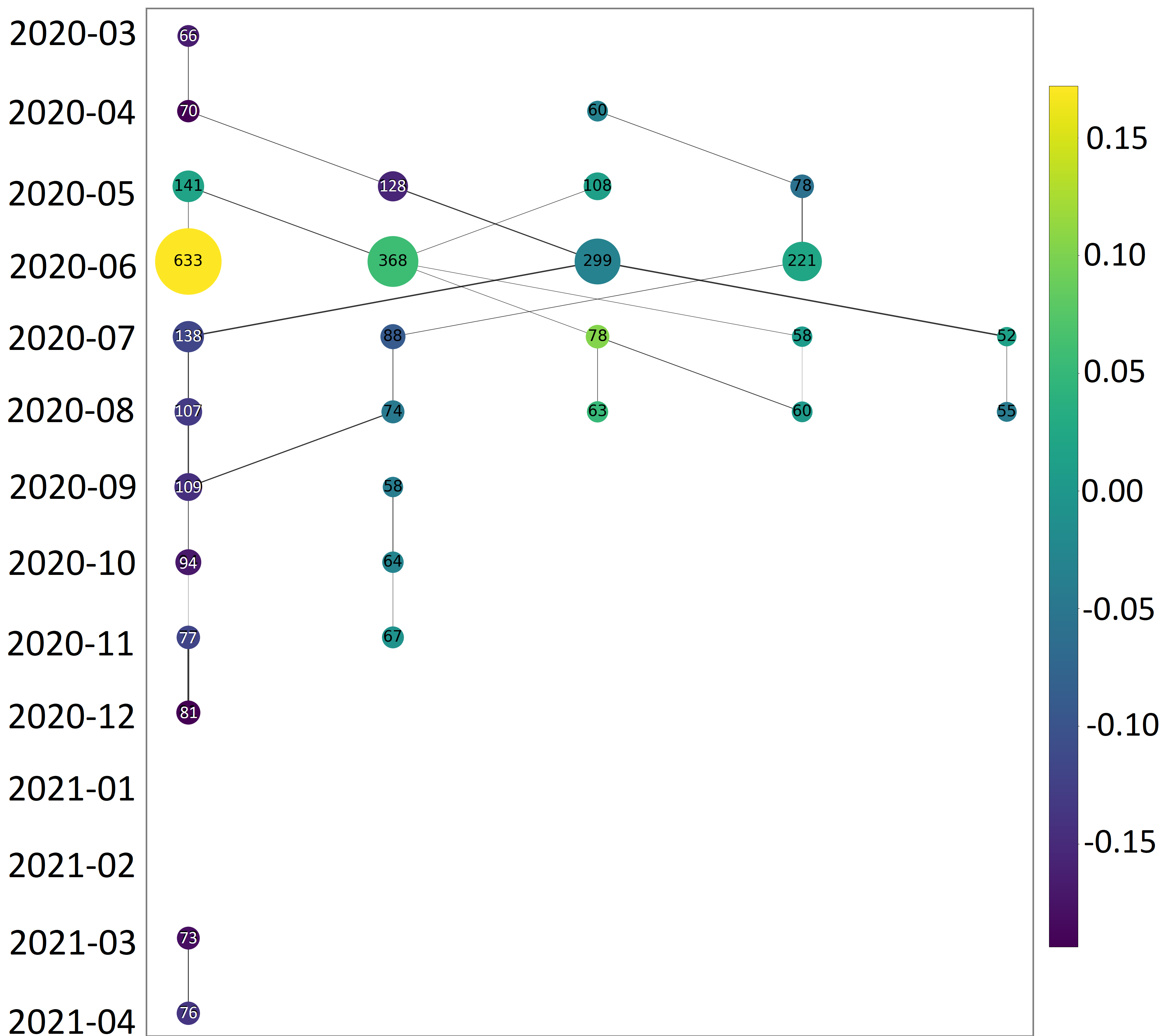}
        \caption{}
        \label{fig:commnunity_users_lgbt}
    \end{subfigure}

    \caption{The evolution of users communities in the a) Black-topic group network and b) LGBTQ+ topic group network ($\gamma = 0.01$). The node size indicates the size of a community (also shown in numbers), the color shows the mean sentiment of the users in the community, and the edge width indicates the number of common users in the connected communities.}
    \label{fig:community_users}
\end{figure*}

To better understand the dynamics of the social network, we examine the evolution of networks involved in the Black and LGBTQ+ topic groups, focusing on the users discussing topics relevant to each topic group over time. Easy extraction of relevant quantities is enabled by the knowledge graph.

We extract user-tweet, tweet-topic, and relevant user-user triples to form respective sub-graphs for the Black and LGBTQ+ topic groups. We then partition undirected sub-graphs into monthly snapshots and apply Louvain community detection \cite{Blondel_2008} to each snapshot. 
We omit the edge weight for tweet-topic (``in\_topic'') relations during community detection to allow better clustering of communities. We record the users within each Louvain community from monthly snapshots. We then use the Jaccard index $J(A, B) = \frac{|A \cap B|}{|A \cup B|}$ to track common users in the communities over time\cite{ogwok2022jaccard}. Thus, users within a community  tweet on similar topics. We connect two user communities $A$ at time $t$ and $B$ at time $t+1$ with an edge if $J(A,B) \geq \gamma$ with $\gamma=0.01$. To reduce the noise, we only consider communities with 100 or more users for the Black topic group and the communities with 50 or more users for the LGBTQ+ topic group.

Figure \ref{fig:community_users} illustrates the dynamics of these user communities over time. Notably, neither of these user communities remains constant over the period from March 2020 to June 2021. Instead, communities appear to emerge and dissolve over time. During May and June 2020, many new, big communities appeared in association with the Black topic group. But these new groups persist for one or two months only. The surge in new communities coincides with George Floyd's murder, capturing a relationship between this public event and the change in online discourse patterns. Moreover, most of these new communities are connected to tweets with positive sentiments, indicating an expression of solidarity with and support for the Black community.

Similarly, the largest communities in the LGBTQ+ topic group appeared in June 2020, corresponding to Pride Month. However, here, not as many new communities emerged and instead communities tended to split into sub-communities over time. This may indicate that discussions within the LGBTQ+ community became more diverse and multifaceted during this period, leading to splitting into sub-topics or sub-interests. Most of these communities expressed positive sentiment during Pride Month, as one may expect, with the largest community having the highest positive sentiment. 
However, throughout the rest of the time period, the average sentiment in the largest LGBTQ+ topic group communities is negative.

Overall, the differing patterns of community evolution between Black and LGBTQ+ topic groups highlight each network's unique characteristics and dynamics. Moreover, in both networks, the change in sentiment and communities over time is closely connected to real-life events, showing that online discussions and sentiment are sensitive to external factors and societal occurrences.

\section{User Network}
  \label{user network}

To analyze the most central users among the Black and LGBTQ+ topic groups, we build Black or LGBTQ+ ``user networks" (similarity networks with users as nodes). These networks examine the similarities between the tweets of active (frequently-posting) users through edges with weight given by similarity in their tweeted topics, using the topic probabilities generated by BERTopic.

\subsection{User Network Construction}

First, we filter the dataset to include 1195 users who tweeted at least ten times in the overall time period. Then, between every pair of users $A,B$ we assign the weight \begin{equation} \label{eq:usernetwork}
    w_{A,B} = \sum_{(t_1, t_2) \in T_A \times T_B} \sum_{\tau \in L} t_1^{(\tau)} \cdot t_2^{(\tau)},
    \end{equation}
\noindent where $T_A$ and $T_B$ are the collections of tweets posted by users $A$ and $B$, respectively, $L$ is a set of selected topics, and $t_i^{(\tau)}$ refers to the probability that the tweet $t_i$ is in topic $\tau$ assigned by BERTopic. 
This formulation defines a ``similarity" between users. Edges with high weights connect users who have similar topic distributions. This produces a complete graph where the weighted degree of a user indicates their overall similarity to other users in the network in terms of \eqref{eq:usernetwork}. Users who tweet on the most common topics tend to have larger weighted degrees as their tweets are more likely to be on the same topics as other users. Summing over every pair of tweets between users also gives weight to users who tweet more frequently. 
We create two different networks by varying $L$: a user network using only Black topics and a user network using only LGBTQ+ topics. 
As such, the top users are different in each network.

\subsection{Exploring Influential Users}

Next, we compute the eigenvector centrality of each user, which measures the importance of the user through the importance of its neighbors. Because edge weights (and consequently centrality) do not account for a user's follower count which can be an important quantity, we introduce influence $I$: \[I = c \cdot \ln(f + 1),\]where $c$ is the user's eigenvector centrality and $f$ their follower count. We take the natural logarithm of the follower count because the distribution of the follower counts is highly skewed. Note that the follower count attribute may include followers (users) that are not in this network or in the dataset, nevertheless it provides a natural estimate of user influence.

We analyze and compare the five most influential users in the Black and LGBTQ+ user networks. Several of their statistics are shown in Table \ref{table:user_stats}. We omit the ID and screen names of the users. We see that the most influential users in the Black user network have more followers and tweet more frequently compared to users in the LGBTQ+ user network. Also, the average sentiments of the top five users in the Black user network are all negative, while the average sentiments of the top five users in the LGBTQ+ user network are all positive.
    
\begin{figure}
    \includegraphics[width=\columnwidth]{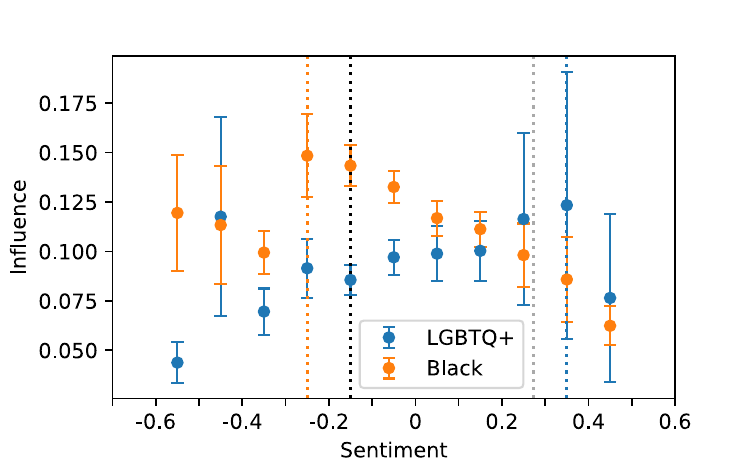}
    \caption{Influence versus sentiment comparison for the 1195 users in the Black and LGBTQ+ user networks. Each data point marks the mean (and standard error) of influence of users whose tweets have sentiments inside the interval on the x-axis, omitting data points that average less than ten users. Blue and orange dotted lines denote the sentiment interval corresponding to the largest influence in the Black and LGBTQ+ user networks, respectively. Black and grey dotted lines denote the average sentiments of the five most influential users in the Black and LGBTQ+ user networks respectively.}
    \label{fig:inflsent}
\end{figure}

Next, for each network, we calculate the mean and standard error of user influences, grouped by the average sentiment of their tweets (Figure \ref{fig:inflsent}). To limit noise, we omit data points that average less than ten users. For the Black user network, we see that the distribution of influence means is skewed to the right with the peak occurring at $-0.25$, indicating that users who tweet with sentiments between $-0.2$ and $-0.3$ are the most influential. The LGBTQ+ user network shows nearly the opposite result with the most influential users tweeting with positive sentiments. The difference in peaks of influence suggests that the most popular users in the Black user network gain traction through posting negatively, or that popularity prompts negativity, as opposed to the trend of positivity among influential users in the LGBTQ+ user network. Furthermore, mean influence has greater standard error among users in the LGBTQ+ user network, perhaps indicating that average sentiment among the LGBTQ+ topic group is more heterogeneous. Interestingly, the mean of the average sentiments of the five most influential users for each network is less strong than each network's influence peak averaged over all active users, implying that the most influential users are not as polarizing as their respective networks overall.
    
\begin{table}
    \centering
    \begin{tabular}{lllll}
      \midrule
      \textbf{User (Black)} & \textbf{Centrality} & \textbf{Followers} & \textbf{\# Tweets} & \textbf{Sentiment} \\
      \midrule
      User 1 & 0.298 & 885 & 417 & -0.223 \\
      User 2 & 0.330 & 163 & 405 & -0.191 \\
      User 3 & 0.224 & 1254 & 371 & -0.080 \\
      User 4 & 0.115 & 17869 & 104 & -0.235 \\
      User 5 & 0.112 & 10411 & 142 & -0.018 \\
      \midrule
      \textbf{User (LGBTQ+)} & \textbf{Centrality} & \textbf{Followers} & \textbf{\# Tweets} & \textbf{Sentiment} \\
      \midrule
      User 1 & 0.244 & 6747 & 69 & 0.018 \\
      User 2 & 0.289 & 606 & 40 & 0.264 \\
      User 3 & 0.206 & 1424 & 52 & 0.127 \\
      User 4 & 0.217 & 592 & 28 & 0.616 \\
      User 5 & 0.199 & 433 & 32 & 0.336 \\
      \bottomrule
    \end{tabular}
    \caption{User statistics for five most influential users in Black and LGBTQ+ user networks.}
    \label{table:user_stats}
\end{table}

\section{Conclusion and Future Work}
\label{conclusion}
In this work, we conduct a multi-faceted quantitative analysis of online sentiment and hate crimes.
Our time series analysis reveals notable temporal correlations between negative sentiment tweets and physical violence against the Black and LGBTQ+ communities and identifies how they are related to large-scale events. The intensity of online sentiment appears to be related to hate crimes, but causality is difficult to infer.

Our topic model-based knowledge graph (KG) provides a framework for studying the link between online social networks and hate crime. Dynamic community detection reflects differences in the discourse on the KG around different topic groups and reveals that the Black topic group communities exhibit a pattern of emergence and dissolution, reflecting the dynamic appearance and disappearance of new topics. On the other hand, the LGBTQ+ topic group communities tend to split, which suggests that discussions within the LGBTQ+ topic group are prone to split into sub-topics or sub-interests.

Finally, we introduce a measure of `influence' by combining network centrality and follower counts of users. Through user network analyses, we find that the most influential users in the Black topic group have higher engagement (tweet frequency and follower count) than the most influential users in the LGBTQ+ topic group. They also post significantly more negatively than users in the LGBTQ+ topic group. Our analysis suggests that the nature of tweets driving engagement is different in different groups---negative sentiments gain traction in the Black topic group, whereas positive sentiments gain traction in the LGBTQ+ topic group. One can leverage this finding to proactively monitor online sentiment trends, informing protective measures for vulnerable groups.

Our research is a robust, large-scale, data-driven analysis, and provides a preliminary analysis on trends in online discourse and hate crimes against marginalized communities. This framework can be extended in many possible subsequent directions, including a more comprehensive analysis, subject to availability of appropriate data, on other areas of the country and other groups, and particularly how separate online communities might affect each others' sentiment over time. Another direction of future work could explore lagged correlations (i.e., where correlations between a pair of time-series where one is time-delayed compared to the other) in order to study time-delayed correlations between online discussions on hate crimes. Lastly, one could enrich the nodes and edges of the knowledge graph with additional features, enabling more detailed analyses. 


\section{Code and Data Availability}
\label{code and data}
A complete online repository of this project's codes is available from \url{https://github.com/jimliu01/kg-twitter-hate-crimes}. For confidentiality, the Twitter dataset is not published. However, tweets are publicly available via Twitter (now known as X) and we list all the processing steps in the paper.

\bibliographystyle{IEEEtran}
\bibliography{references}

\end{document}